\documentclass[twocolumn,amssymb,nobibnotes, aps, pra,showpacs,superscriptaddress]{revtex4-1}

\usepackage{amsmath}
\usepackage{verbatim}
\usepackage{epsfig}
\usepackage{subfigure}
\usepackage[breaklinks]{hyperref}
\usepackage{graphicx}  
\usepackage{dcolumn}   
\usepackage{bm}        
\usepackage{amssymb}   
\usepackage{color}
\usepackage{adjustbox}
\newcommand{\ket}[1]{\left|{#1}\right\rangle}

\newcommand{\modu}[1]{\left|{#1}\right|}

\newcommand{\bea}{\begin{eqnarray}}
\newcommand{\eea}{\end{eqnarray}}
\begin{document}

\title{Positive energy density leads to no squeezing}%
\author{S. Kannan}
\altaffiliation[Present address: ]{ISRO Inertial Systems Unit, Thiruvananthapuram, 695 013, India.}
\email{kns.phys@gmail.com}
\affiliation{Department of Physics, Indian Institute of Space Science and Technology, Thiruvananthapuram, 695 547, India.}

\author{C. Sudheesh}%
\email{sudheesh@iist.ac.in}
\affiliation{Department of Physics, Indian Institute of Space Science and Technology, Thiruvananthapuram, 695 547, India.}
\date{\today}
\begin{abstract}
We consider  two kinds of superpositions  of squeezed states of light. In the case of superpositions of first kind, the squeezing and all higher order squeezing  vanishes.  However, in the case of the second kind, it is possible to achieve a maximum amount of squeezing by adjusting the parameters in the superposition.  The emergence and vanishing of squeezing for the superposition states are explained on the basis of expectation values of the energy density. We show that   expectation values of energy density of quantum states which show no squeezing will be always positive and that  of squeezed states will be negative for some values of spacetime-dependent phase.
\end{abstract}
\maketitle
\section{Introduction}\label{sec1}
Squeezed states are a class of nonclassical states of light which have less fluctuations in one quadrature phase than a coherent state at the expense of increased fluctuations in the other \cite{mollow1967quantum,stoler1970equivalence,*stoler1971equivalence,yuen1976two,walls1983squeezed}. These states offer intriguing applications such as in optical communication systems \cite{yuen1978optical,shapiro1979optical}, quantum sensing \cite{lawrie2019quantum}, noise-free amplification \cite{majeed1991effect}, gravitational wave detection \cite{caves1981quantum,abadie2011gravitational,aasi2013enhanced} etc. Recently the increase in sensitivity using squeezed light in LIGO and VIRGO detectors has been reported \cite{tse2019quantum,acernese2019increasing}. A rapid increase in the use of squeezed states in the continuous variable quantum information processing is also seen \cite{braunstein2005quantum}. Engineering interactions between electric dipoles utilizing the antisqueezed  quadrature of a squeezed vacuum state has been predicted recently. This does not require any photonic structures \cite{zeytinouglu2017engineering}. As many such theoretical and practical implications exist, investigations of squeezing properties of various quantum states is still a central topic in quantum optics.

As pointed out by Dirac, the linear superposition principle is one of the most fundamental features of quantum mechanics. It has been shown that quantum superposition can give rise to various nonclassical effects such as squeezing, higher-order squeezing \cite{hong1985higher}, sub-Poissonian statistics, etc. Superposition of two coherent states \cite{buvzek1992superpositions}, superposition leading to even and odd squeezed coherent states \cite{gerry1993generation,hach1994generation}, superposition of squeezed coherent states \cite{zhu1993nonclassical}, binomial state \cite{stoler1985binomial,joshi1989effects}, intermediate number-phase state \cite{baseia1995intermediate} and generalized superposition of two squeezed states \cite{barbosa2000generalized} are some examples for the  nonclassical states.

Despite the fact that there are  a lot of studies in the literature which talk about squeezing due to superposition of states but there is no work  which clearly states  what type of superpositions leads to squeezing and its connection with any other physical parameter of the system. However, there are some attempts has been made  in the literature  \cite{buvzek1992superpositions,schleich1991nonclassical} to address  the aforementioned questions.  In this paper, we  analyze different types of superposition states  to find   what  type  of superpositions give rise to squeezing, and to find a  relation between squeezing and  expectation value of the energy density.   Another motivation to do this study is to devise a method to get higher  or no squeezing when we add two squeezed states. These two analyses  will definitely help researchers working in quantum communication and quantum sensing  using squeezed light to choose proper squeezed states for their study. For this purpose,  we  consider the superposition of various squeezed states such as the squeezed vacuum states \cite{yuen1976two}, photon-added coherent states \cite{agarwal1991nonclassical} etc. We investigate both the nonclassical and statistical features of the states and identify the role of superposition in modifying the quantum fluctuations in the field quadratures.

The paper is organized as follows: in Section \ref{sec2} we study the squeezing and higher-order squeezing of superposition of various squeezed states, vanishing and enhancing of squeezing for different superposition is seen. In Section \ref{sec3} expectation value of energy density is used to explain the squeezing properties shown by the superposition states and Section \ref{sec4} is devoted to conclusion.

\section{Superposition of squeezed states and its squeezing}\label{sec2}

Squeezing can be quantified by analyzing the uncertainties in the quadrature operators. For a single-mode scenario, the quadrature operators are defined as
\begin{equation}\label{quad1}
	\hat{X}=(\hat{a}+\hat{a}^{\dagger})/2,\textit{ } \hat{P}=(\hat{a}-\hat{a}^{\dagger})/2i.
\end{equation}
They satisfy the commutation relation
\begin{equation}\label{commu}
	[\hat{X},\hat{P}]=i/2
\end{equation}
and hence the uncertainty relation
\begin{equation}\label{uncert}
	\langle(\Delta \hat{X})^{2}\rangle\langle(\Delta \hat{P})^{2}\rangle\geq 1/16.
\end{equation}  
(We have set $\hbar=1$.) A given state is said to be quadrature squeezed if the variance in one of the quadrature falls below the vacuum value:
\begin{equation}\label{squee}
	\langle(\Delta \hat{X})^{2}\rangle< 1/4 \text{ or } \langle(\Delta \hat{P})^{2}\rangle< 1/4.
\end{equation}
Here we only consider the superposition of the same class of quadrature squeezed states. We can classify the superposition of these types of states into two: (1)  all the states in the superposition have the same squeezing value, (2)  a generalized superposition where the states have different squeezing values. The first kind of superposition can be expressed as
\begin{equation}\label{sup1}
	\ket{\Psi}_{l}=N_{l}\sum_{j=0}^{l-1}\ket{\Phi e^{i2\pi j/l}},
\end{equation} 
where $ N_{l} $ is the normalization constant and $ \ket{\Phi} $ can be any squeezed state of the same class. Here, it is always possible to find an observable which  can turn one of the  states in the superposition  to another one in the superposition. The second kind, the generalized superposition is of the form:
\begin{equation}\label{supg}
	\ket{\Psi}_{gen}=N_{gen}\sum_{j=0}^{l-1}a_{j}\ket{\Phi_{j}}.
\end{equation}
Here $ N_{gen} $ denotes the normalization factor, $ \ket{\Phi_{j}} $ represents squeezed states of the same class, and $ a_{j} $ the corresponding weight factors.

Let us now analyze the quadrature variance of superpositions of some important squeezed states.
\subsection{Superpositions of the first kind}\label{sec2a}
\subsubsection{Squeezed vacuum state}
The squeezed vacuum state $ \ket{\xi} $ is generated by the action of squeezing operator  $\hat{S}(\xi) $ \cite{stoler1970equivalence,yuen1976two,walls1983squeezed} on the vacuum state: 
 \begin{equation}\label{svs}
 \ket{\xi}=\hat{S}(\xi)\ket{0},
 \end{equation} where $ \xi = re^{i\theta}$ with real $ r $ (squeezing parameter) and $ 0\leq\theta\leq 2\pi $. Experimental realization of this state was achieved way back in 1985-1986 \cite{slusher1985observation,wu1986generation}. In the recent years, great progress in the generation of this state have been achieved \cite{andersen201630,otterpohl2019squeezed,gao2019generation,zhao2020near} that could significantly improve quantum information applications.

In the Fock basis, $ \ket{\xi} $ can be expanded as
 \begin{equation}\label{eq3}
\ket{\xi}=\dfrac{1}{\sqrt{\textrm{cosh(r)}}}\sum_{m=0}^{\infty}\left(-e^{i\theta}\textrm{tanh(r)}\right)^{m}\dfrac{\sqrt{(2m)!}}{2^{m}m!}\ket{2m}.
\end{equation}
For this state, the variance in the both quadratures come out to be \cite{gerry2005introductory}
 \begin{equation}\label{var1}
	\langle(\Delta \hat{X})^{2}\rangle=\dfrac{1}{4}e^{-2r} \text{ and } \langle(\Delta \hat{P})^{2}\rangle=\dfrac{1}{4}e^{2r}, 
\end{equation} 
and,  hence,  there is a squeezing in the $\hat{X}$-quadrature. Now, consider a superposition of $ l $ squeezed vacuum state $ \ket{\zeta}_{l} $ of the first kind:
\begin{equation}\label{supsvs}
\ket{\zeta}_{l}=N_{s}\sum_{j=0}^{l-1}\ket{\xi e^{i2\pi j/l}}, 
\end{equation}
where $ N_{l} $ is the normalization constant and is equal to
\begin{equation}\label{mormal1}
N_{l}=\left(\sum_{m=0}^{\infty}\dfrac{(2lm)!}{2^{2lm}(lm)!^{2}}\left(\textrm{tanh(r)}\right)^{2lm}\right)^{-1/2}.
\end{equation}
In the Fock basis,
\begin{equation}\label{supsvsfock}
\ket{\zeta}_{l}=N_{l}\sum_{m=0}^{\infty}\left(-e^{i\theta}\textrm{tanh(r)}\right)^{lm}\dfrac{\sqrt{(2lm)!}}{2^{lm}(lm)!}\ket{2lm}.
\end{equation}
The above superposed states can be produced  by utilizing the techniques in \cite{o2019hybrid} . We find that  for the states $ \ket{\zeta}_{l\text{ } \geq\text{ } 2} $,
 \begin{equation}\label{positn}
	\langle\hat{a}\rangle=\langle\hat{a}^{\dag}\rangle=0
\end{equation}
 and
\begin{equation}\label{secondord}
	\langle\hat{a}^{2}\rangle=\langle\hat{a}^{\dag 2}\rangle=0.
\end{equation}
Since the above above expectation values are zero, we get 
\begin{equation}\label{varia}
	\hspace{-2mm}\langle(\Delta
	\hat{X})^{2}\rangle\hspace{-1mm}=\hspace{-1mm}\langle(\Delta
	\hat{P})^{2}\rangle\hspace{-1mm}=\hspace{-1mm}\dfrac{1}{4}+\dfrac{N_{l}^{2}}{2}\sum_{m=0}^{\infty}\left(\text{tanh(r)}\right)^{2lm}\dfrac{2lm(2lm)!}{2^{2lm}(lm)!^{2}}.
\end{equation}
Since the RHS of the above equation is always greater than $1/4$, there is no squeezing for $ \ket{\zeta}_{l\text{ } \geq\text{ } 2} $ in both the quadratures.
It is  one of the important results of this paper.  To check whether this is true for any superposition of  squeezed states of first kind, we examine superpositions of other types of squeezed states in the following subsections.  
%
%

\subsubsection{Photon-added coherent state}
$m$-photon-added coherent states (m-PACS) were introduced as  the states which are intermediate between the Fock states and the coherent states. The Fock state representation  of $m$-PACS is  
\begin{equation}\label{pacs}
\ket{\alpha,m}=\dfrac{e^{-{\modu{\alpha}^{2}}/2}}{\sqrt{L_{m}(-\modu{\alpha}^{2})m!}}\sum_{n=0}^{\infty}\frac{\alpha^{n}}{n!}\sqrt{(n+m)!}\ket{n+m},
\end{equation}
where $ L_{m} $ is the Laguerre polynomial of order $ m $, $ \alpha $ (coherent parameter) is a complex number and $ m $ is an integer.
These states show rich  nonclassical properties and  exhibit quadrature squeezing for a wide range of $\alpha$ values  \cite{agarwal1991nonclassical}.  Experimental generation of the single photon-added coherent state has been achieved \cite{zavatta2004quantum,*zavatta2005single}, and there also exists schemes for the generation of higher-order photon-added coherent states \cite{sivakumar2011photon,shringarpure2019generating}.  
Consider the  $l$ superposition of the first kind of $m$-PACS:
\begin{equation}\label{supPACS}
  \ket{\Gamma,m}_{l}=N_{m}\sum_{j=0}^{l-1}\ket{\alpha e^{i2\pi j/l},m},
 \end{equation}
where $N_m$ is the normalization constant. Since we are examining  only  superposition of squeezed states,   we consider  range of $ \alpha $ for which the state $ \ket{\alpha,m} $ shows quadrature squeezing.  We then calculate the quadrature variance for the state $ \ket{\Gamma,m}_{l} $ for different $ m $ and  $ l $ values. Our calculations show that  the quadrature squeezing is  also absent for all $m$ values for the state $ \ket{\Gamma,m}_{l\text{ } \geq\text{ } 2} $. 
We confirm our result also for superposition of  two-mode squeezed states.

\subsubsection{Two-mode squeezed vacuum state}
Quadrature squeezing is observed in multi-mode scenarios also. The most common one is the two-mode squeezed vacuum state $ \ket{\xi}_{2}  $. It is generated by the action of a two-mode squeeze operator $ \hat{S_{2}}(\xi) $ \cite{caves1982quantum,*caves1985new} on the vacuum state:
\begin{equation}\label{twosvs}
	\begin{aligned}
		\ket{\xi}_{2}&=\hat{S_{2}}(\xi)\ket{0,0}\\&=\dfrac{1}{\textrm{cosh(r)}}\sum_{n=0}^{\infty}\left(-e^{i\theta}\textrm{tanh(r)}\right)^{n}\ket{n,n},
	\end{aligned}
\end{equation}
where $ \xi = re^{i\theta}$.   Since there is a  correlation between the modes,  squeezing  exist in the effective  quadrature operators:
\begin{equation}\label{twoqo}
\hat{X_{1}}=(\hat{a}+\hat{a}^{\dagger}+\hat{b}+\hat{b}^{\dagger})/{2^{3/2}},
\text{ }\hat{X_{2}}=(\hat{a}-\hat{a}^{\dagger}+\hat{b}-\hat{b}^{\dagger})/{i2^{3/2}}.
\end{equation}
Here also we consider the first kind of superposition  and variance in the above quadratures operators   are analyzed. For the two superposition state \begin{equation}\label{eq20}
N\left(\ket{\xi}_{2}+\ket{-\xi}_{2}\right)=\dfrac{2N}{\textrm{cosh(r)}}\sum_{n=0}^{\infty}e^{in\theta}(\textrm{tanh(r)})^{2n}\ket{2n,2n},
\end{equation} with normalization constant $ N $, the variance in the quadratures $ \hat{X_{1}} $ and $ \hat{X_{2}} $ comes out to be the same and is equal to \begin{equation}\label{eq21}
\Delta \hat{X}_{1}^{2}=\Delta \hat{X}_{2}^{2}=\dfrac{1}{4}\left(1+\left(\dfrac{2N}{\text{cosh(r)}}\right)^{2}\sum_{n=0}^{\infty}4n\left(\text{tanh(r)}\right)^{4n}\right).
\end{equation} 
So there is no squeezing in the superposition state and we confirm also the  absence of quadrature squeezing  for all higher numbers  (i.e., $l>2$) of superpositions. In the above three cases, we have shown that  no first-order squeezing exists when we take superposition of the  first kind with squeezed states. In the next section, we extend our analysis for higher-order squeezing of superposition of squeezed states. 

We have also calculated both the   Hong and Mandel  \cite{hong1985higher}  and  Hillery type \cite{hillery1987amplitude}  higher order squeezing  for  the superposition of squeezing states.  Our calculations shows that there is
 no higher-order squeezing exists for the superposition of the states which we have considered earlier.
 In this section,  we have shown  that superposition of squeezed states can lead to complete vanishing of squeezing properties.  This is quite unexpected, since usually the superposition of wavepackets give rise to squeezing properties (e.g. even-cat state \cite{dodonov1974even}).

\subsection{Generalized superposition}\label{2.C}
We have defined the generalized superposition state $ \ket{\Psi}_{gen} $ in Eq. \ref{supg} as the superposition of squeezed states with different squeezing value and weight factors. Here, we analyze the variance and higher-order moments in the quadrature of  $ \ket{\Psi}_{gen} $ for squeezed vacuum states:
\begin{equation}\label{svsgen}
\ket{\Psi}_{gen}=N_{gen}\sum_{j=0}^{l-1}a_{j}\ket{\xi_{j}},
\end{equation}
where $ \xi_{j}=r_{j}e^{i\theta_{j}}$, $ a_{j} $'s are the weight factors and the normalization constant  
\begin{equation}\label{Nsvsg}
N_{gen}=\left(\sum_{i=1}^{l}\sum_{j=1}^{l}a_{i}a_{j}^{*}\dfrac{1}{\sqrt{\text{cosh}(r_{i}-r_{j})}}\right)^{-1/2}.
\end{equation}
The squeezing and statistical properties of the two superposition cases have been discussed by Barbosa et al. in \cite{barbosa2000generalized}. Here, we  also find squeezing properties of more than two superpositions and in turn find  suitable weight factors  to achieve maximum quadrature squeezing.
The variance in the  $ \hat{X} $ quadrature for the state given in Eq. \ref{svsgen}  comes out to be
\begin{equation}\label{eq31}
\begin{aligned}
&\langle(\Delta\hat{X})^{2}\rangle= { }{ } \dfrac{1}{4}+\dfrac{\text{N}_{gen}^{2}}{4}\Bigg[\sum_{i=1}^{\emph{l}}\sum_{j=1}^{\emph{l}}\sum_{n=0}^{\infty}a_{i}a_{j}\dfrac{\left(\text{tanh}(r_{i})\text{tanh}(r_{j})\right)^{n}}{\sqrt{\text{cosh}(r_{i})\text{cosh}(r_{j})}}\\&\times\dfrac{(2n)!}{2^{2n}(n!)^{2}}\Bigg(4n-(2n+1)\Big(\text{tanh}(r_{i})+\text{tanh}(r_{j})\Big)\Bigg)\Bigg].
\end{aligned}
\end{equation}
Without loss of generality, we take real values for  $ \xi_{j} $'s and $ a_{j} $'s. Using a minimization algorithm \cite{scipy}, the weight factors ($ a_{i}$'s) are calculated to achieve  the highest squeezing. For fixed squeezing parameters, the weight factors are computed such that the variance  attains the minimum possible value.  We found that the state $ \ket{\Psi}_{gen} $  can have quadrature variance less than $ \dfrac{1}{4}e^{-2R} $, where $ R=\text{max}(r_{j}) $.   The minimum value of variance obtained for different numbers of superpositions along with the corresponding weight factor is tabulated  below.
	\begin{center}
	\begin{table}[h]
		\begin{center}
		    \caption{}
			\label{table1}
			\begin{tabular}{|l|c|c|c|}\toprule
				\text{S/N} & \text{ Squeezing parameters} & \text{ Variance}   & \text{ Weight factors} \\
				& ($ r_{j} $) & $ \langle(\Delta\hat{X})^{2}\rangle $ & ($ a_{j} $)\\
				\hline
				1 & \begin{tabular}{@{}c@{}}
					$ \emph{l}=1 $,\\ $ r_{1}=1 $
				\end{tabular} & 0.0338 & $ a_{1}=1 $\\
				\hline
				2 &\begin{tabular}{@{}c@{}}
					$ \emph{l}=2 $,\\ $ r_{1}=0.5 $, $ r_{2}=1 $
				\end{tabular} & 0.0268 & \begin{tabular}{@{}c@{}}
					$ a_{1}=-0.32678 $,\\ $ a_{2}=1 $
				\end{tabular}  \\
				\hline
				3 & \begin{tabular}{@{}c@{}}
					$ \emph{l}=3 $,\\ $ r_{1}=0.5 $, $ r_{2}=0.8 $, $ r_{3}=1 $
				\end{tabular} & 0.0188 &  \begin{tabular}{@{}c@{}}
					$ a_{1}=-0.2317 $,\\ $ a_{2}=-1.0103 $,\\ $ a_{3}=1 $
				\end{tabular}  \\
				\hline
				4 & \begin{tabular}{@{}c@{}c@{}}
					$ \emph{l}=4 $,\\ $ r_{1}=0.5 $, $ r_{2}=0.7 $,\\ $ r_{3}=0.8 $, $ r_{4}=1 $
				\end{tabular} & 0.0151 &  \begin{tabular}{@{}c@{}}
					$ a_{1}=-0.3464 $,\\ $ a_{2}=2.4050$,\\ $ a_{3}=-2.9717 $,\\ $ a_{4}=1 $
				\end{tabular} \\
				\botrule
			\end{tabular}
		\end{center}
	\end{table}
\end{center}
It is clear from the table that it is possible to increase the squeezing  by increasing the superposition  in the state. We have also studied the higher-order squeezing the state and found that  the state exhibits  higher-order squeezing. For  example,  the state $ \ket{\Psi}_{gen} $ with $ l=3 $, the fourth and sixth-order squeezing comes out to be $ \langle (\Delta X)^{4}\rangle=0.0068 $ and $ \langle (\Delta X)^{6}\rangle=0.1857 $, respectively.

In this section, we have shown that the two different types of superpositions  give two contrasting results:   complete vanishing of squeezing to all order and  a greater amount of squeezing by adjusting the weight factors in the superposition. This enables us to control the squeezing by adjusting the parameters in the superpositions which will have application in quantum information processing. More important than this, the above result leads  us to find a connection between energy density and squeezing which we describe in detail in the next section.
\section{Squeezing and energy density}\label{sec3}
An obvious question that arises now is, why is the squeezing in some superposition state vanishing? Or why is it increasing in certain cases? It is  well known that some superposition of coherent state, such as even cat state \cite{dodonov1974even} and Yurke-Stoler states \cite{yurke1986generating} exhibit quadrature squeezing while coherent state doesn't have such a nonclassical property. In this section, we explain these properties based on the expectation value  of the energy density of these quantum states.\\
It is known that all forms of classical matter have non-negative energy density but this is not the case for a general quantum state. A superposition of number eigenstates can give a negative expectation value of energy density in certain spacetime regions due to coherence effect.  To find the expectation value of the energy density, we calculate the expectation value of the stress-energy tensor or energy-momentum tensor \cite{charles1973gravitation}. Stress-energy tensor describes the density and flux of energy and momentum in spacetime. In order to calculate this expectation value for the stress-energy tensor, the method described in reference \cite{kuo1993semiclassical} is followed. Consider a massless and minimally coupled scalar field. Its Lagrangian density is given by
 \begin{equation}\label{lagra}
\mathcal{L}=\dfrac{1}{2}\eta_{\mu\nu}(\partial^{\mu}\phi)(\partial^{\nu}\phi),
\end{equation}
 where the spacelike metric $ \eta_{\mu\nu}= $  diag[-1,1,1,1] and $ \phi $ is the quantum field operator which satisfies the dynamical equation
\begin{equation}\label{d'alam}
\left(-\dfrac{\partial^{2}}{\partial\text{t}^{2}}+\nabla^{2}\right)\phi(x)\equiv \partial_{\mu}\partial^{\mu}\phi(x)=0.
\end{equation}
We have taken $ c=\hslash=1 $.The quantum field operator $ \phi $ can also be expanded in mode functions as
\begin{equation}\label{phi}
\phi=\sum_{k}(a_{k}f_{k}+a^{\dagger}_{k}f^{*}_{k}),
\end{equation}
where $ a_{k} $ satisfy the usual bosonic commutation relation and the mode function $ f_{k} $ is given by
\begin{equation}\label{fk}
f_{k}=(2L^{3}\omega)^{-1/2}e^{i\textbf{k.x}-i\omega t}.
\end{equation}
 Here we have assumed periodic boundary condition in a three-dimensional box of side L and $ \omega $ is the frequency of the mode and \textbf{k} the wave number. For our purpose,  we consider the stress tensor  corresponding   to the single   mode excitation:
\begin{equation}\label{stress}
T_{\mu\nu}=(\partial_{\mu}\phi)(\partial_{\nu}\phi)-\dfrac{1}{2}\eta_{\mu\nu}(\partial_{\sigma}\phi)(\partial^{\sigma}\phi).
\end{equation}
We calculate the normal ordered (same as subtracting the vacuum expectation value) expectation value of this stress tensor for the quantum states of our interest. We will be only looking at  the expectation value  of the temporal component ($  \langle:T_{00}:\rangle $). For coherent state $ \ket{\alpha} $, this normal ordered expectation value comes out to be \begin{equation}\label{eq37}
\langle:T_{00}:\rangle=2K_{00}\alpha^{2}(1-\text{cos}(2\theta)),
\end{equation}
where $K_{00}=\dfrac{(k_{0})^{2}}{2\omega L^{3}} $ and the spacetime-dependent phase $\theta=k_{\rho}x^{\rho}$. Without loss of generality, the above expression is derived for real values of $\alpha$. It is easy to see that for coherent states, $\langle:T_{00}:\rangle\geq 0$ and increases with increase in $\alpha$ value for all values of $\theta$. For  even cat state 
\begin{equation}\label{eq38}
 \ket{\psi}=\dfrac{2}{\sqrt{2(1+e^{-2\modu{\alpha}^{2}})}}\sum_{n=0}^{\infty}\dfrac{\alpha^{2n}}{\sqrt{(2n)!}}\ket{2n},
\end{equation}
 with real $ \alpha $, the energy density comes out to be
 \begin{equation}\label{eq39}
 \langle:T_{00}:\rangle=2K_{00}\alpha^{2}\text{tanh}(\alpha^{2})\Big(1-\text{coth}(\alpha^{2})\text{cos}(2\theta)\Big). 
 \end{equation} 
 For fixed  values of $ \alpha $, the energy density becomes negative for some values of $\theta$ in $[0,2\pi]$.  
 Thus for the even cat state, which shows quadrature squeezing, energy density becomes negative for certain values of spacetime-dependent  phase $ \theta $. In contrast, we have found earlier that a coherent state, which shows no squeezing, never gives negative energy density for any values of $\theta$.  To check that whether the above result is a general result applicable to all quantum states, we carry out the above analysis to other important class of states and superposition states considered in 
 the Section  \ref{sec2}.

 The energy density for odd cat state, which shows no squeezing, is always positive. For the squeezed vacuum state defined in Eq. \ref{eq3}the average energy density 
  \begin{equation}\label{eq40}
 \langle:T_{00}:\rangle=2K_{00}\text{sinh(r)}\Big(\text{cosh(r)}\text{cos}(2\theta)+\text{sinh(r)}\Big).
 \end{equation}
 Again, we have considered real values for $\xi$  $(=r) $ without loss of generality. 
Here, the energy density  can go below zero for certain values of $ \theta $ for a given $r$ and for larger values of $r$ the  energy density tends to zero. This can be associated with larger oscillations in the photon number distribution \cite{schleich1987oscillations1}. For the superposition state of first kind given in Eq. \ref{supsvs} with $\emph{l}=2$, which shows no squeezing, the average energy density 
\begin{equation}\label{eq41}
 \langle:T_{00}:\rangle=\dfrac{16K_{00}\text{sech(r)}}{\left(1+\sqrt{\text{sech(2r)}}\right)}\Big(\sum_{n=0}^{\infty}\dfrac{n(4n)!}{(2n)!^{2}}(\text{tanh(r)}/2)^{2n}\Big)
\end{equation}
 is always greater than zero. Higher order superposition will also give rise to a positive average energy density. Our results are also verified  for PACSs (Eq. \ref{pacs}) and their superpositions. Finally for the generalized states given in TABLE. \ref{table1}, the expectation value for the stress energy tensor is found to have negative values.
 
 One of the  important results from this section can be stated as follows:  The  expectation value of energy density of quantum states which show no squeezing will be always positive and that  of squeezed states will have negative for some values of spacetime-dependent phase.
 Our analysis also gives the reason for  absence of squeezing in superposition states of first kind and presence of squeezing in superposition  of states of second kind in terms of expectation value of the energy density.

 \section{Conclusion}\label{sec4}
We have studied the squeezing properties of  arbitrary numbers of superpositions of  various squeezed states such as squeezed vacuum state, photon-added coherent states and two mode squeezed vacuum state.  We have considered two kinds of superpositions: first kind and second kind. In the case of first kind, the superpositions of squeezed states doesn't show both squeezing and higher-order squeezing of all orders. This is found to be true for any state which has quadrature squeezing  and  multi-mode squeezed states.   However,  in the case of the superpositions of second kind (also called as generalized superpositions), it has been shown that  the superposition states show large amounts of quadrature and higher-order squeezing. This is achieved by  choosing the proper weight factors in the superpositions; this method also enables us to have a control over the amount of squeezing produced. 

 The vanishing and appearance of squeezing in superposition state is explained on the basis of expectation values of energy density or stress-energy tensor. States with squeezing are shown to have a negative expectation value for the stress-energy tensor for some values of spacetime-dependent  phase. In the case of states with no squeezing, the expectation values of energy density is always positive.  We have found that in the case of superposition of squeezed states of  first kind, the expectation values of energy density is  always positive  and in the case of second kind,  the average value is negative for some values of  spacetime-dependent  phase.  
 \bibliography{reference1}
\end{document}